\documentstyle[graphicx, times, amsmath, amssymb]{mn2e}
\title[The Suzaku View of NGC 7213]
{Evidence for a Truncated Accretion Disc in the Low Luminosity Seyfert Galaxy, NGC 7213?}

\newcommand{\arcm}{\hbox{$^\prime$}}

\newcommand{\ltsima}{$\; \buildrel < \over \sim \;$}
\newcommand{\simlt}{\lower.5ex\hbox{\ltsima}} % < over ~
\newcommand{\gtsima}{$\; \buildrel > \over \sim \;$}
\newcommand{\simgt}{\lower.5ex\hbox{\gtsima}} % > over ~

\author[A. Lobban et al.] {
A.P. Lobban$^1$, J.N. Reeves$^1$, D. Porquet$^2$, V. Braito$^3$, A. Markowitz$^4$, L. Miller$^5$ and T.J. Turner$^6$\\
$^1$Astrophysics Group, School of Physical and Geographical Sciences, Keele 
University, Keele, Staffordshire, ST5 8EH, UK\\
$^2$Observatoire astronomique de Strasbourg, Universit\'{e} Louis-Pasteur, CNRS, INSU, 11 rue de l'Universit\'{e}, 67000 Strasbourg, France\\
$^3$Department of Physics and Astronomy, University of Leicester, University Road, Leicester, LE1 7RH, UK\\
$^4$Center for Astrophysics and Space Sciences, University of California, San Diego, M.C. 0424, La Jolla, CA, 92093-0424, USA\\
$^5$Department of Physics, University of Oxford, Denys Wilkinson Building, Keble Road, Oxford, OX1 3RH, UK\\
$^6$Department of Physics, University of Maryland Baltimore County, Baltimore, MD 21250 and Astrophysics Science Division, NASA/GSFC, Greenbelt, MD 20771, USA
}

\date{Accepted by MNRAS on 4 June 2010}
\pagerange{\pageref{firstpage}--\pageref{lastpage}} \pubyear{2010}

\begin{document}

\def\aap{A\&A}
\def\apj{ApJ}
\def\apjl{ApJ}
\def\mnras{MNRAS}

\maketitle

\label{firstpage}

\begin{abstract}

We present the broad-band 0.6--150\,keV 
{\sl Suzaku} and {\sl Swift} BAT spectra of the low luminosity Seyfert galaxy, NGC 7213. 
The time-averaged continuum emission is well fitted by a single power-law of 
photon index $\Gamma = 1.75$ 
and from consideration of the {\sl Fermi} flux limit we constrain the high energy cutoff to 
be 350\,keV \textless $E_{\rm cut}$ \textless 25\,MeV. 
Line emission from both near-neutral iron K$\alpha$ at 6.39\,keV 
and highly ionised iron, from Fe\,\textsc{xxv} and Fe\,\textsc{xxvi}, is strongly detected 
in the {\sl Suzaku} spectrum, 
further confirming the results of previous observations with {\sl Chandra} and 
{\sl XMM-Newton}. We find the centroid energies for the emission from Fe\,\textsc{xxv} 
and Fe\,\textsc{xxvi} to be 6.60\,keV and 6.95\,keV respectively, with the latter 
appearing to be resolved in the {\sl Suzaku} spectrum. 
From modelling, we show that the Fe\,\textsc{xxv} and Fe\,\textsc{xxvi} 
emission can result from a highly photo-ionised plasma, with a column 
density of $N_{\rm H} \sim 3\times10^{23}$\,cm$^{-2}$.
A Compton reflection component, e.g., originating from an optically-thick accretion 
disc or a Compton-thick torus, appears either very weak or absent in this 
AGN, subtending $<1$\,sr to the X-ray source, consistent with previous findings.
Indeed the absence of Compton reflection from either 
neutral or ionised material coupled
with the lack of any relativistic Fe K signatures in the spectrum suggests that an inner, 
optically-thick accretion
disc is absent in this source. 
Instead, the accretion disc
could be truncated with the inner regions perhaps replaced by a Compton-thin 
Radiatively Inefficient Accretion Flow (RIAF). 
Thus, the Fe\,\textsc{xxv} and Fe\,\textsc{xxvi} emission could both originate in ionised 
material perhaps at the transition region between the hot, inner flow and the cold, 
truncated accretion disc on the order of $10^{3} - 10^{4}$ gravitational radii from the 
black hole. The origin for the unresolved neutral 
Fe\,K$\alpha$ emission is then likely to be 
further out, perhaps originating in the optical Broad Line Region or a Compton-thin 
pc-scale torus.

\end{abstract}


\section{Introduction}

NGC 7213 is a nearby low-luminosity AGN ($z=0.005839$), often
classified as an intermediate between a Seyfert 1 and a LINER (Low-Ionisation Nuclear Emission-line Region) galaxy due to its optical spectrum (Filippenko \& Halpern 1984). Its X-ray
spectral properties also appear to lie between those of weak
AGN (e.g., M81) and 'classical' higher-luminosity
broad-line Seyferts.
The ultraviolet (UV) flux measured by Wu, Boggess \& Gull (1983) was higher
than would be expected from an extrapolation of the optical flux,
indicating that NGC 7213 may have a Big Blue Bump (BBB), although weak
compared to most Seyferts.  
This object has a high black hole mass of about 10$^{8}$M$_{\odot}$
as estimated from the stellar velocity dispersion (Nelson \& Whittle 1995; Woo \& Urry 2002) and a low bolometric luminosity
  ($L_{\rm bol}$) of about 9 $\times$ 10$^{42}$ erg\,s$^{-1}$ (Starling et al. 2005). \\

NGC 7213 has a very low accretion rate of only $0.07\%$ $L_{\rm
  Edd}$, a value which is intermediate between those usually found in local Type 1
Seyfert Galaxies (e.g., Padovani \& Rafanelli 1988; Wandel 1999; Page 2001) and LINERs
(e.g., Ho 1999).
Interestingly, this is much less than the predicted 2\% $L_{\rm Edd}$
``critical'' rate whereby the high/soft state in black hole X-ray binaries can
be observed (Maccarone 2003). Furthermore, this object exhibits another
interesting characteristic since it is part of a class of Seyfert
galaxies which have radio properties that are intermediate between those of radio-loud
and radio-quiet active galaxies (e.g., Blank, Harnett \& Jones 2005 and reference
therein). It is therefore conceivable that NGC 7213 is an analogue of the
Galactic low/hard state sources. \\

A simultaneous {\sl XMM-Newton} (net EPIC-pn exposure $\sim$30\,ks) and {\sl BeppoSAX}
observation in May 2001 revealed further peculiar characteristics of NGC 7213. The low S/N RGS (Reflection Grating Spectrometer) spectrum showed the presence of
several weak emission features with no absorption lines (Starling et al. 2005) contrary to what is usually found in Seyfert Type 1 Galaxies. Moreover, the emission lines
appeared to be the signature of a collisionally-ionised thermal plasma
($k_{\rm B}T \sim 0.18$\,keV), while in Seyfert Galaxies, only emission and/or absorption
lines from a photo-ionised warm absorber / emitter have before been observed
(e.g., NGC 3783, Kaspi et al. 2001; NGC 4151, Schurch \& Warwick 2002; NGC 1068, Kinkhabwala et al. 2002; Brinkman et al. 2002; Mrk 3, Pounds \& Page 2005).
Such emission lines from collisionally-ionised thermal plasma have, however,
been observed in the soft X-ray spectra of LINERs, such as
M81 (Page et al. 2003). 
Interestingly, no significant Compton reflection was observed in this simultaneous {\sl XMM-Newton} and {\sl BeppoSAX} observation (reflection component: $R = \Omega / 2\pi < 0.2$, where a value $R = 1$ corresponds to reflection 
off material subtending $2 \pi$ sr; Bianchi et al. 2003),
though the presence of a significant
Fe\,K complex could be explained by three narrow
emission lines: neutral iron at around 6.40\,keV with an EW of $\sim$80\,eV, Fe\,\textsc{xxv} at around 6.66\,keV
and Fe\,\textsc{xxvi} at about 6.94\,keV (see also Starling et al. 2005). \\

Bianchi et al. (2003) deduced from the absence of the
reflection component that the neutral iron K$\alpha$ emission line is most likely
produced in a Compton-thin torus or the Broad Line Region (BLR).
Indeed, according to Matt, Perola \& Pirlo (1991) and George \& Fabian (1991), a line
with an EW of $\sim$80\,eV would require a reflection component of
about $R\sim 0.6$, a larger value than that found for this object ($R < 0.2$).
Furthermore, Bianchi et al. (2008) reported the data analysis of a long
{\sl Chandra}/HETG observation of NGC 7213 finding that the neutral iron K$\alpha$
line is resolved with a FWHM value of 2\,400$^{+1100}_{-600}$\,km\,s$^{-1}$, fully consistent with the H$\alpha$ line width
(2\,640$^{+110}_{-90}$\,km\,s$^{-1}$) measured with the ESO/NTT telescope.
They therefore inferred that the neutral Fe\,K line seen in this object
originates in the Compton-thin BLR explaining the lack of evidence for Compton
reflection. They also
confirmed the presence of two ionised iron lines at $\sim$6.72\,keV
and $\sim$6.99\,keV which they associate most probably with the
resonance transition of the Fe\,\textsc{xxv} triplet and the Ly$\alpha$ transition of Fe\,\textsc{xxvi}, 
respectively. Comparing the line energies found with their rest-frame values, a
blue-shift of about 900\,km\,s$^{-1}$ is inferred.
If the dominant line at $\sim$6.72\,keV  is indeed the
resonance line of the Fe\,\textsc{xxv} triplet, then this means that the line
may be associated with a collisionally-ionised thermal plasma (Porquet \& Dubau 2000; Bautista \& Kallman 2000).\\

Here we report on a 90\,ks {\sl Suzaku} (Mitsuda et al. 2007) observation of NGC 7213. 
The overall goal is to understand accretion in an AGN at low rates compared to Eddington, 
through a high signal-to-noise, broad-band observation of this source.
Specifically, the objectives are to parameterise the iron line complex with an analysis of the XIS (X-ray Imaging Spectrometer) spectra in order to constrain the individual properties of the lines before considering the complete broad-band spectra from 0.6--150\,keV with a combined analysis of the {\sl Suzaku} XIS and HXD (Hard X-ray Detector) data with that from the time-averaged {\sl Swift} BAT 22-month survey (see Section 3). The aims are to obtain better constraints on the origin of both the iron line complex and any observed soft excess whilst simultaneously testing for the presence (or absence) of a Compton reflection component (see Section 4).

\section{Suzaku Analysis and Data Reduction}

\subsection{Suzaku Analysis}

NGC 7213 was observed by {\sl Suzaku} on October 22 2006 with
a total net exposure of 90\,ks.
In this paper we discuss data taken from the 4 XIS
(Koyama et al. 2007) CCDs and the PIN diodes 
of the HXD (Takahashi et al. 2007). \\

Events files from version 2.0.6.13 of the {\sl Suzaku} pipeline processing were used.
All events files were screened within \textsc{xselect} to exclude data
taken within the SAA (South Atlantic Anomaly) as well as excluding
data with an Earth elevation angle (ELV) $<5$ degrees. Data taken with
Earth day-time elevation angles (DYE\_ELV) less than 20 degrees were
also excluded. A cut-off rigidity (COR) criteria of $>6$\,GeV/c for
the XIS was applied. Only good events with grades 0,2,3,4 and 6 were
used, while hot and flickering pixels were removed from the XIS images
using the \textsc{cleansis} script. Time intervals affected by
telemetry saturation were also removed. \\

Subsequently, source spectra from the XIS CCDs were extracted from circular regions of 2.3\arcm\ radius centered on the source, in the on-axis
XIS nominal pointing position. Background spectra were
extracted from 2.3\arcm\ circles offset from the source region, avoiding
the calibration sources on the corners of the CCD chips. XIS response
files (RMFs) and ancillary response files (ARFs) were generated using
the \textsc{xisrmfgen} and \textsc{xissimarfgen} \textsc{ftools} respectively including
correction for the hydrocarbon contamination on the optical blocking
filter (Ishisaki et al. 2007).  A net XIS source exposure of 90.7\,ks
was obtained for each of the 4 XIS chips.  The 3 front-illuminated
XIS chips (XIS 0,2,3; hereafter XIS--FI) are predominantly used in this paper as they have the greatest 
sensitivity at iron K. These chips were found to produce consistent spectra within
the statistical errors, so the spectra and responses were combined to
maximise signal to noise. The net source count rate for the 3 XIS
combined was $1.585\pm0.002$\,counts\,s$^{-1}$ per XIS, with the background rate only
0.7\% of the source rate. This count rate corresponds to an observed flux of $2.46\times10^{-11}$\,erg\,cm$^{-2}$\,s$^{-1}$ and a luminosity of $1.85\times10^{42}$\,erg\,s$^{-1}$ over the 2--10\,keV range. The XIS source spectrum was binned at the HWHM resolution of the detector due to the high photon statistics. This enabled the use of $\chi^{2}$
minimisation as there were \textgreater 50 counts per resolution bin.  
Errors are quoted to 90\% confidence for 1 parameter (i.e., $\Delta\chi^{2}=2.7$) 
unless otherwise stated.

\subsection{HXD Reduction}

As NGC 7213 is below the detection threshold of the HXD/GSO, we used data from the HXD/PIN only, where this object is detected at the 13$\sigma$ level relative to the background.
The source spectrum was extracted from the cleaned HXD/PIN events
files and processed with the screening criteria described above. The
HXD/PIN instrumental background spectrum was generated from a ``tuned'' time
dependent model provided by the HXD instrument team (Fukazawa et al. 2009). Both
the source and background spectra were made with identical GTIs (Good
Time Intervals) and the source exposure was corrected for detector
deadtime (which is $\approx6.7\%$). A detailed description of the PIN
detector deadtime is given in Kokubun et al. (2007). The net exposure
time of the PIN source spectrum was 84.3\,ks after deadtime
correction. Note that the background spectral model was generated with
$ 10 \times$ the actual background count rate in order to minimise the
photon noise on the background; this has been accounted for by
increasing the effective exposure time of the background spectra by a
factor of $ \times 10$. The HXD/PIN response file dated 2008/01/29 (epoch 3) for the XIS nominal position was used for these spectral
fits. \\

In addition, a spectrum of the cosmic X-ray background (CXB) (Boldt 1987; Gruber et al. 1999) was also simulated with the HXD/PIN. The
form of the CXB was taken as $9.0\times10^{-9}(E/3$\,keV$)^{-0.29}$exp$(-E/40$\,keV$)$\,erg\,cm$^{-2}$\,s$^{-1}$\,sr$^{-1}$\,keV$^{-1}$. When normalised to the field
of view of the
  HXD/PIN instrument, the effective flux of the CXB component is $8.49\times10^{-12}$\,erg\,cm$^{-2}$\,s$^{-1}$
  in the 15--50\,keV band corresponding to a count rate of $\sim$0.017\,counts\,s$^{-1}$. The net flux of NGC 7213 measured by
  the HXD over the same band is $3.58\times10^{-11}$\,erg\,cm$^{-2}$\,s$^{-1}$, i.e., the CXB component
  represents 24\% of the net source flux measured by the
  HXD/PIN. Note that there may be some uncertainty in the absolute flux
  level of the CXB component measured between missions; for instance, Churazov et al. (2007) find the CXB normalisation from INTEGRAL to
  be about 10\% higher than measured by Gruber et al. (1999) from the
  HEAO-1 data. However, a factor of $\pm 10\%$ uncertainty in the CXB normalisation would result in a $\pm 2.4\%$ uncertainty in the HXD flux for NGC 7213, which is 
well within the statistical uncertainty of the HXD/PIN observations. After background subtraction (including both the instrumental and CXB components), the resulting net PIN source count rate from 15--50\,keV was $0.062\pm0.002$\,counts\,s$^{-1}$ corresponding to a 15--50\,keV flux of $3.58\times10^{-11}$\,erg\,cm$^{-2}$\,s$^{-1}$. Note that the total background count rate was $\sim$0.350\,counts\,s$^{-1}$ (15--50\,keV) with a typical $1 \sigma$ systematic uncertainty of $\pm 1.3\%$. \\

We used 0.6--10\,keV data in both the XIS--FI and XIS--BI spectra. We ignored the 1.7--1.9\,keV band in the co-added
FI spectrum and the BI spectrum due to uncertainties
in calibration associated with the instrumental Si\,K edge. In all
fits, we included a constant multiplicative factor to account for relative instrument
normalisations. We allowed the relative XIS--BI/XIS--FI normalisation to vary,
but best-fit values were always within 1\% of each other. \\

A visual analysis of the lightcurves was undertaken to determine whether any detailed timing analysis was necessary. It can be seen from Figure 1 that the amplitude of the XIS--FI lightcurve varies only by a factor of $\sim$0.1 throughout the entire observation indicating little intrinsic variability below 10\,keV. From Figure 2 it can be seen that the HXD/PIN lightcurve, too, shows little evidence of any substantial variability in the hard X-ray band. Therefore, due to the lack of any strong evidence of short-timescale spectral variability, we proceed to consider the time-averaged spectrum (Section 3).

\begin{figure}
\begin{center}
\rotatebox{-90}{\includegraphics[width=5cm]{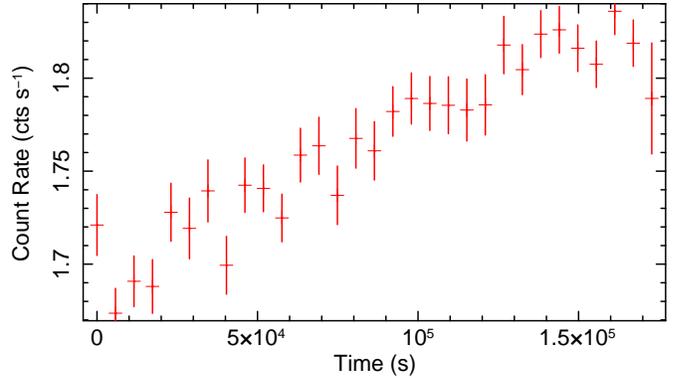}}
\end{center}
\caption{XIS--FI background-subtracted lightcurve of NGC 7213 from 0.5--10\,keV in 5760\,s orbital bins.}
\end{figure}

\begin{figure}
\begin{center}
\rotatebox{-90}{\includegraphics[width=5cm]{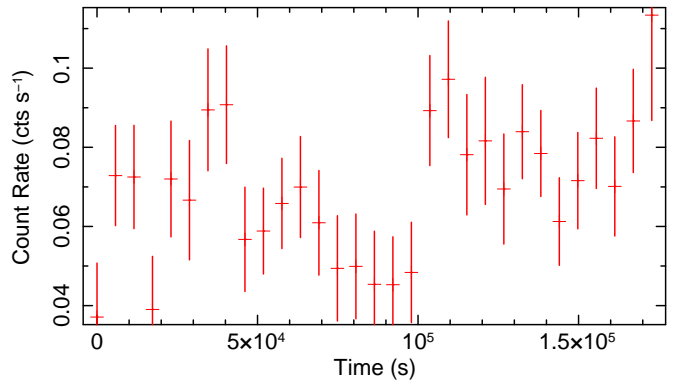}}
\end{center}
\caption{Net background-subtracted HXD/PIN lightcurve of NGC 7213 from 15--50\,keV in 5760\,s orbital bins.}
\end{figure}

\section[]{Spectral analysis}\label{sec:spectra}

The {\sc xspec v11.3} software package (Arnaud 1996) was used for spectral
analysis of the background-subtracted spectrum.
In all fits, we included the Galactic column density
($N^{\rm Gal}_{\rm H}$ = 1.1 $\times 10^{20}$\,cm$^{-2}$,
 obtained from the \textsc{ftool nh} using the compilations of Dickey \& Lockman 1990 and Kalberla et al. 2005) 
and used the cross-sections for X-ray absorption by
the interstellar medium from Morrison \& McCammon (1983).
Note that all fit parameters are given in the rest frame of the galaxy, assuming a distance of 25\,Mpc to NGC 7213 (Mould et al. 2000).
 Abundances are those of Anders \& Grevesse (1989) unless otherwise stated. \\

The cross-normalisation between the HXD/PIN and XIS detectors was accounted for by the addition of a fixed constant component at a value of 1.16 for the XIS nominal pointing position, a value derived using {\sl Suzaku} observations of the Crab (Ishida, Suzuki \& Someya 2007\footnote{ftp://legacy.gsfc.nasa.gov/suzaku/doc/xrt/suzakumemo-2007-11.pdf}).

\begin{figure}
\begin{center}
\rotatebox{-90}{\includegraphics[width=6cm]{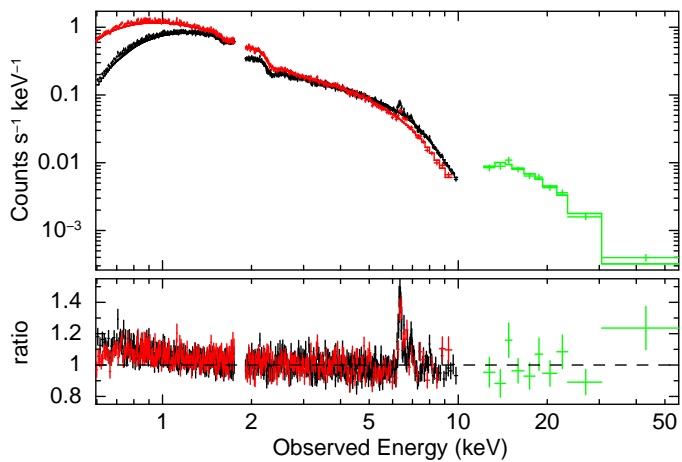}}
\end{center}
\caption{The {\sl Suzaku} spectra of NGC 7213 (in the observed frame) showing the XIS--FI (black), XIS--BI (red) and HXD/PIN (green).
An absorbed (Galactic column density) power-law has been fit to the broad-band spectrum. 
A significant positive residual is observed in the Fe\,K complex
energy range, as well as a weak excess in the $<1$\,keV energy
range.}
\end{figure}

\begin{figure}
\begin{center}
\rotatebox{-90}{\includegraphics[width=6cm]{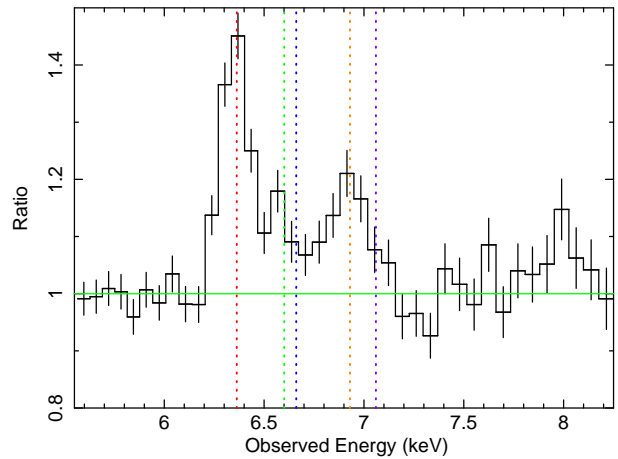}}
\end{center}
\caption{A plot of the ratio of the residuals for XIS--FI of the Fe line complex compared to the power-law continuum. The spectra are binned relative to the half-width at half-maximum of the detector resolution. The vertical dotted lines show the expected line energies of, from left to right, Fe\,K$\alpha$, Fe\,\textsc{xxv} forbidden, Fe\,\textsc{xxv} $1s$--$2p$ resonance, Fe\,\textsc{xxvi} $1s$--$2p$\,Ly$\alpha$ and Fe\,K$\beta$ in the observed frame.}
\end{figure}

\begin{table*}
\centering
\begin{tabular}{l c c c c}
\hline\hline
Model Component & Fit Parameter & Value & $\chi^2/{\rm dof}$ & $\Delta \chi^{2}$
 \\ [0.5ex]
\hline
1. Power-law Continuum$^{a}$ & $\Gamma$ & $1.75\pm0.02$ \\ & Normalisa
tion & $6.25^{+0.02}_{-0.06}$ \\ & $F_{\rm 2.5-10\,keV}$ & $2.18\times10^{-11}$ \\
2. Galactic Absorption$^{b}$ & $N_{\rm H}$ & $1.10\times10^{20}$ & $739/316$ \\
3. Fe\,K$\alpha$ Line$^{c}$ & $E_{\rm line}$ & $6.39\pm0
.01$ & $422/313$ & $316.8$ \\ & $\sigma$ & $<4.30\times10^{-2}$ \\ & EW & $83.1^{+11.0}_{-10.7}$ \\ & FWHM & $<4600$ \\ & Line Flux & $2.18^{+0.
28}_{-0.29}$ \\
4. H-like Line$^{c}$ & $E_{\rm line}$ & $6.95\pm0.03$ & $365/310$ & $57.1$
\\ & $\sigma$ & $0.10^{+0.05}_{-0.04}$ \\ & EW & $62.3^{+16.0}_{-14.2}$ \\ & FWHM & $10000^{+5000}_{-4000}$ \\ & Line Flux & $1.40^{+0.36}_{-0.32}$ \\
5. 6.60\,keV Line$^{c}$ & $E_{\rm line}$ & $6.60\pm0.03$ & $332/307$ & $32.8$
 \\ & $\sigma$ & $<0.30$ \\ & EW & $24.4\pm8.0$ \\ & FWHM & $<32000$ \\ & L
ine Flux & $0.67\pm0.22$ \\
6. 8.00\,keV Line$^{c}$ & $E_{\rm line}$ & $8.00^{+0.10}_{-0.14}$ & $322/304$ & $10.0$ \\ & $\
\sigma$ & $<0.28$ \\ & EW & $45.9^{+33.5}_{-27.7}$ \\ & FWHM & $<25000$ \\ &
 Line Flux & $0.78^{+0.57}_{-0.47}$ \\
7. Fe\,K$\beta$ Line$^{c}$ & $E_{\rm line}$ & $7.06$ & $322/304$ & $0.5$ \\ & $\sigma$
& $<4.30\times10^{-2}$ \\ & EW & $< 29.3$ \\ & FWHM & $<4600$ \\ & Line Flux
& $< 0.67$ \\
8. Fit Statistics$^{d}$ & $\chi^{2}/{\rm dof}$ & $322/304$ \\ & Null Proba
bility & $0.36$ \\
\hline
\end{tabular}
\caption{Spectral Parameters to {\sl Suzaku} XIS in the Fe\,K band.
$^{a}$ $\Gamma$, photon index; Normalisation in units $\times10^{-3}$\,photons\,cm$^{-2}$\,s$^{-1}$; $F_{\rm 2.5-10\,keV}$, absorbed continuum flux from $2.5$--$10$\,keV in units ergs\,cm$^{-2}$\,s$^{-1}$. 
$^{b}$ Local Galactic absorption (at $z=0$), units cm$^{-2}$. 
$^{c}$ $E_{\rm line}$, line energy in units keV; $\sigma$, 1$\sigma$ line width in units keV; EW, equivalent width in units eV; FWHM, full width at half maximum in units km\,s$^{-1}$; Line Flux in units $\times10^{-5}$\,photons\,cm$^{-2}$\,s$^{-1}$. 
$^{d}$ Reduced chi-squared ($\chi^{2}/dof$) and null hypothesis probability for spectral fit.}
\end{table*}

\subsection{The Fe\,K line profile}

The X-ray spectrum was initially analysed in the 0.6--50\,keV band using both the XIS--FI and HXD/PIN data.  A power-law with Galactic absorption of column density $N_{\rm H} = 1.1 \times10^{20}$\,cm$^{-2}$ was fitted to the data revealing a slight soft excess at energies $<1$\,keV as shown in Figure 3.  The XIS\, 1 data were included in this fit to check for consistency.  For clarity, the HXD data were binned to $10\sigma$ per spectral bin relative to the background.  The hard X-ray data are seen to extrapolate quite well to the XIS data with very few residuals in excess of the power-law continuum indicating little or no reflection component ($R\sim0.2$, see Section 3.2).  Line emission is clearly present with a strong but seemingly relatively narrow Fe K$\alpha$ line at $\sim$6.4\,keV.  As the XIS\, 1 has a much lower $S/N$ ratio at higher energies above $2$ keV, these data were initially excluded from the Fe\,K line analysis leaving the XIS--FI to be analysed from 2.5--10\,keV.  The HXD data were also initially excluded as the lack of Compton reflection suggested that the Fe\,K emission lines could be modelled independently.  The HXD and XIS\,1 data are re-included in the broad-band fits in Section 3.2. \\

A simple power-law model of $\Gamma=1.75\pm0.02$ with Galactic absorption evidently resulted in a poor fit ($\chi^{2}/dof=739/316$) highlighted by a low null hypothesis probability of $6.67\times10^{-36}$.  A plot of the ratio of the residuals with respect to the power-law continuum from 5.0--8.5\,keV (Figure 4) clearly shows X-ray line emission which requires modelling; the most apparent being the Fe\,K$\alpha$ line from near-neutral material at $6.39\pm0.01$\,keV with an intrinsic width of $\sigma<4.30 \times 10^{-2}$\,keV, equivalent width of EW = $83.1^{+11.0}_{-10.7}$\,eV and an observed flux of $F_{\rm K\alpha}=2.18^{+0.28}_{-0.29}\times10^{-5}$\,photons\,cm$^{-2}$\,s$^{-1}$.  Adding this line improves the fit significantly with a value of $\Delta\chi^{2}=316.8$ for 3 parameters of interest.  However, even upon modelling the strong Fe\,K$\alpha$ line at 6.39\,keV, the fit remains unacceptable (null probability $= 3.70\times10^{-5}$) with further residuals still present between 6.5--7.0 keV indicating K-shell emission from ionised Fe. \\

Further Gaussian lines were added to fit other prominent emission lines starting with the $1s$--$2p$ doublet from hydrogen-like iron (Fe\,\textsc{xxvi}) at $6.95\pm0.03$\,keV which corresponds to a value of $\Delta\chi^{2}=57.1$ for an additional 3 parameters of interest. Unlike the $6.39$\,keV line, this line appears to be resolved compared to the detector resolution with an intrinsic width of $\sigma=0.10^{+0.05}_{-0.04}$\,keV (FWHM $\sim 10\,000$\,km\,s$^{-1}$), an equivalent width of EW = $62.3^{+16.0}_{-14.2}$\,eV and an observed flux of $F_{\rm line}=1.40^{+0.36}_{-0.32}\times10^{-5}$\,photons\,cm$^{-2}$\,s$^{-1}$.  A third narrow component at a line energy of $6.60\pm0.03$\,keV was also modelled, improving the fit by a factor of $\Delta\chi^{2}=32.8$ for a further 3 parameters of interest. A line energy vs. line flux contour plot (Figure 5) shows that this line energy can be rejected at the 99.9\% confidence level (for two interesting parameters) as being associated with the resonance line of helium-like iron (Fe\,\textsc{xxv}) at 6.700\,keV ($\Delta\chi^{2}\sim14$) and is just acceptable (rejected at only 90\% confidence) as the forbidden line at 6.637\,keV (also see Section 3.3.2).\footnote{An analysis of the spectrum produced by the XIS Fe-55 calibration source, which produces emission lines from Mn\,K$\alpha$ and Mn\,K$\beta$, shows that the absolute XIS energy scale is accurate to within $\pm 10$\,eV.} The fact that this indicates that the line at 6.60\,keV is more consistent with the forbidden transition of Fe\,\textsc{xxv} is an interesting discovery since the resonance line is expected to dominate over the forbidden line in a collisionally-ionised plasma (Porquet \& Dubau 2000; Bautista \& Kallman 2000). Further discussion regarding the origin of this emission can be found in Section 4.2. \\

A weak, narrow component appeared to remain in the residuals at a line energy of $8.00^{+0.10}_{-0.14}$\,keV with an intrinsic width of $\sigma<0.28$\,keV.  However the detection is more marginal; 
adding this extra line component only improved the fit by $\Delta\chi^{2}=10.0$ for 3 parameters. This putative line could possibly be associated with the $1s$--$3p$ transitions of Fe\,\textsc{xxv} or Fe\,\textsc{xxvi} (corresponding to rest energies of $\sim$7.88 and $\sim$8.25\,keV respectively) or alternatively, it could be due to the $1s$--$2p$ transition of H-like nickel (corresponding to a rest energy of $\sim$8.10\,keV). Finally, no neutral K$\beta$ emission was apparent but was still modelled for consistency at a fixed line energy of $7.06$\,keV, with an intrinsic width, $\sigma$, tied to that of the corresponding K$\alpha$ line. The upper limit on the flux of $F_{\rm K\beta}<0.67\times10^{-5}$\,photons\,cm$^{-2}$\,s$^{-1}$ then provided an upper limit on the K$\beta$/K$\alpha$ flux ratio of 0.35. A value of $\Delta\chi^{2}=0.5$ revealed that the the K$\beta$ line is consistent with the fit to the data but in this instance is not required. Hereafter we include the K$\beta$ line fixed at 13\% of the K$\alpha$ flux in all subsequent fits to remain consistent with the theoretical flux ratio for neutral iron (Kaastra \& Mewe 1993). \\

We did also attempt to model the neutral Fe\,K$\alpha$ emission with a \textsc{diskline} component (Fabian et al. 1989) to test for the presence of any broad, relativistic emission from the inner regions of the accretion disc. We fixed the width of the original Gaussian at 6.39\,keV to be narrow ($\sigma = 10$\,eV) in order to model emission from distant material and introduced a \textsc{diskline} component to the model with the centroid energy fixed at 6.39\,keV and the emissivity index fixed at a standard value of $q = 3$. We also fixed the inner and outer radii of the emission at 6 and 400\,R$_{\rm g}$ from the black hole respectively (where 6\,R$_{\rm g}$ corresponds to the innermost stable orbit for a Schwarzschild black hole) and the inclination angle of the source at $\theta = 30^{\circ}$. Upon fitting, this returned a value for the flux of the line of $F_{\rm diskline} < 7.64 \times 10^{-6}$\,photons\,cm$^{-2}$\,s$^{-1}$ which corresponds to a 90\% upper limit on the equivalent width of the line of EW \textless 26.8\,eV. This tight constraint appears to exclude the presence of any Fe\,K emission from the inner accretion disc. We note that no other significant emission or absorption lines are found in the Fe\,K band.  The values of all of the final parameters and fit statistics are noted in Table 1.

\begin{figure}
\begin{center}
\rotatebox{-90}{\includegraphics[width=6cm]{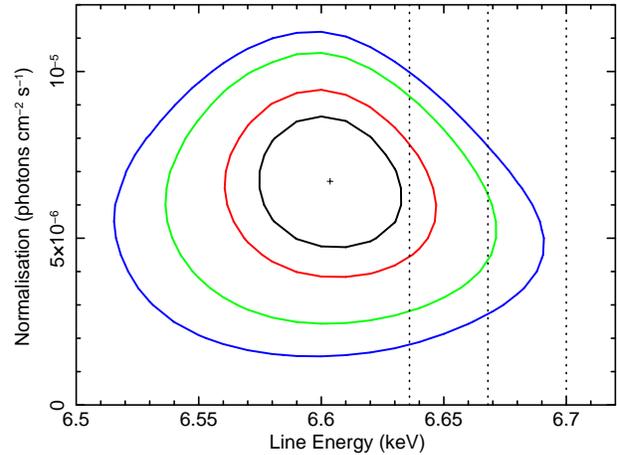}}
\end{center}
\caption{A two dimensional contour plot between line energy and normalisation of the 6.60\,keV line at the 68\%, 90\%, 99\% and 99.9\% confidence intervals (inner to outer contours respectively) for 2 parameters of interest. The vertical dashed lines show the rest energies of forbidden, intercombination and resonance transitions of the Fe\,\textsc{xxv} triplet at 6.637, 6.668 and 6.700\,keV respectively.}
\end{figure}

\begin{figure}
\begin{center}
\rotatebox{-90}{\includegraphics[width=6cm]{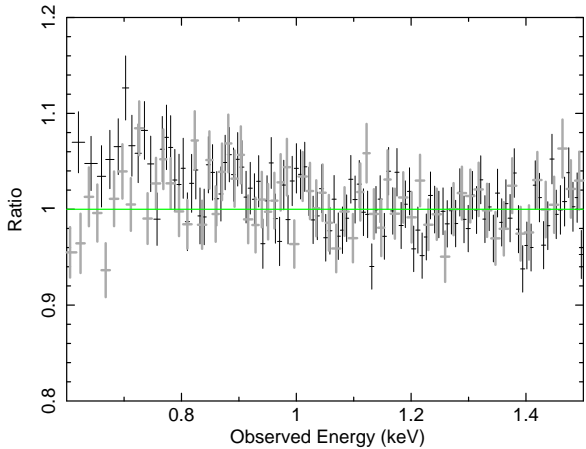}}
\end{center}
\caption{A ratio plot showing the divergence of the XIS 1 detector data (light grey) with that of the XIS 0,\,2 and 3 detectors (black) in the soft X-ray regime (see Section 3.2).}
\end{figure}

\subsection{The Broad-Band Spectrum}

The first stage of the broad-band spectral analysis was to model the spectra above 10\,keV by including the {\sl Suzaku} HXD/PIN data. To check for consistency, the 22-month time-averaged {\sl Swift} BAT spectra (14--150\,keV) were also included (Tueller et al. 2009). This provided an extension of the spectrum far beyond the high energy limit of the HXD PIN. We restricted the energy range of the HXD from 15--50\,keV and applied a constant multiplicative factor of 1.16 to account for the cross-normalisation at the XIS nominal pointing position. The constant in front of the BAT data was allowed to be free as the 14--150\,keV flux over the 22-month period (2004/12/15--2006/10/27) may have varied (time-averaged flux $F_{14-150} = 5.31\times10^{-11}$\,erg\,cm$^{-2}$\,s$^{-1}$). This provided a cross-normalisation factor of $0.75\pm0.11$ for the BAT compared to the {\sl Suzaku} XIS. The power-law component was replaced by an exponential cutoff power-law and we note that no cutoff energy is required in these data. We constrain the 90\% lower limit on $E_{\rm cut} > 350$\,keV. A simultaneous {\sl XMM-Newton} and {\sl BeppoSAX} PDS observation published by Bianchi et al. (2004) required a high energy cutoff with a value of $E_{\rm cut} = 90^{+50}_{-20}$\,keV. However, fixing the cutoff energy at 90\,keV in our {\sl Suzaku}+BAT spectrum results in a worse fit with $\Delta \chi^{2} = 35$ supportive of the notion that the cutoff appears to be at significantly higher energy in these data. \\ 

The residuals in the hard X-ray regime show very little excess flux above 10\,keV when modelled with a power-law indicating a lack of a Compton reflection component. To test for this, we included the \textsc{pexrav} model (Magdziarz \& Zdziarski 1995) which is an additive component incorporating the reflected continuum from a neutral slab. We tied the photon index of the power-law incident upon 
the reflector to that of the power-law continuum and fixed the elemental abundances to Solar (Anders \& Grevesse 1989). We also fixed the cosine of the inclination angle of the source to 0.87 and tied the folding energy to the cutoff energy of the power-law at $\sim$1000\,keV, consistent with no cut-off as above. The inclusion of the \textsc{pexrav} component resulted in a reflection scaling factor value of $R=0.18^{+0.23}_{-0.14}$, consistent with the {\sl XMM-Newton}/{\sl BeppoSAX} analysis of Bianchi et al. (2004) who find $R < 0.19$. The inclusion of this component corresponds to $\Delta\chi^{2}\approx5.0$ suggesting that this component is only marginally required. This resulted in a fit statistic $\chi^{2}$/$d.o.f.$$=486/446$, acceptable with a null hypothesis probability of 0.10. \\

The next step was to model the entire broad-band spectra by including the XIS data below 2.5\,keV. As the signal-to-noise ratio decreases at lower energies, the spectra were only included down to 0.6\,keV for each XIS. The data were also ignored from 1.7--1.9\,keV so as not to include the silicon absorption edge due to the detectors. The inclusion of these data resulted in a slightly worse fit with a null probability of $2.98 \times 10^{-3}$. Residuals were observed at energies $<2$\,keV hinting at the presence of a weak soft excess (e.g., Figure 3). Upon closer inspection of these residuals it was noted that the XIS 1 (BI) detector data slightly diverged with that of the remaining XIS detectors (Figure 6), even when the photon index of the power-law continuum was allowed to vary between detectors.  As the XIS 0,\,2 and 3 (FI) were all self-consistent, this divergence was possibly caused by calibration effects around the oxygen K detector edge. To account for this, the data from the XIS 1 detector were ignored below 0.7\,keV.\\ 

In an attempt to then model the observed soft excess, we added a \textsc{mekal} thermal plasma component incorporating the emission spectra from a hot diffuse gas (Starling et al. 2005). It is worth noting that a featureless blackbody component models the soft excess with an equally good fit as there are no strong lines. However, this model was not considered any further as the lack of evidence for a strong Big Blue Bump (Wu et al. 1983) suggested little thermal emission directly from the disc (Starling et al. 2005). The addition of the \textsc{mekal} component gave a best-fitting thermal plasma temperature of $k_{\rm B}T=0.27^{+0.05}_{-0.04}$\,keV and resulted in an overall better fit with $\chi^{2}$/$d.o.f.$$=1104/1022$ compared to $1151/1024$ before the \textsc{mekal} component was added\footnote{Modelling the soft excess with an \textsc{apec} component was also considered (http://hea-www.harvard.edu/APEC/; Smith et al. 2001) although at the {\sl Suzaku} resolution here, the fit was identical to that obtained with the \textsc{mekal} thermal plasma.}. The luminosity of the \textsc{mekal} component was calculated at $L=2.14\times10^{40}$\,erg\,s$^{-1}$ corresponding to only 1\% of the the total luminosity in the 0.5--10\,keV band. As no further significant residuals were observed in the spectra (Figure 7), this became our accepted broad-band model, the final parameters of which are summarised in Table 2 (note that the full broad-band model from 0.6--150\,keV also resulted in a tighter constraint on the reflection scaling factor, $R$, the best-fit value of which is shown in Table 2). A plot of the relative unfolded model contributions is shown in Figure 8.

\begin{figure}
\begin{center}
\rotatebox{-90}{\includegraphics[width=6cm]{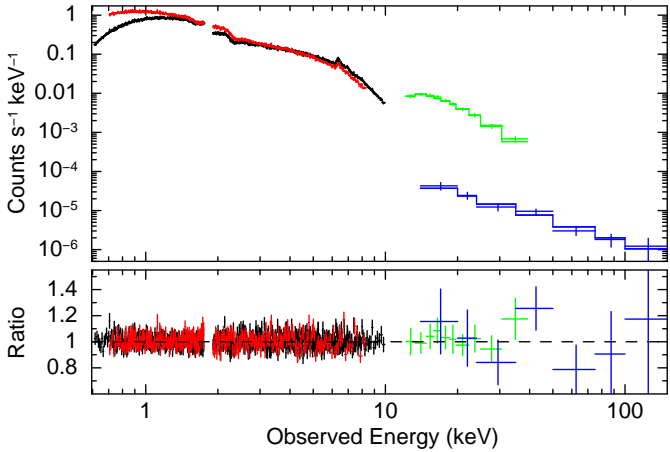}}
\end{center}
\caption{The broad-band spectra of NGC 7213 (in the observed frame) with the inclusion of the {\sl Swift} BAT spectra (blue). A plot of the ratio of the residuals to the model described in Section 3.2 and Table\,2 is shown in the lower panel. No significant residuals are present.}
\end{figure}

\begin{figure}
\begin{center}
\rotatebox{-90}{\includegraphics[width=6cm]{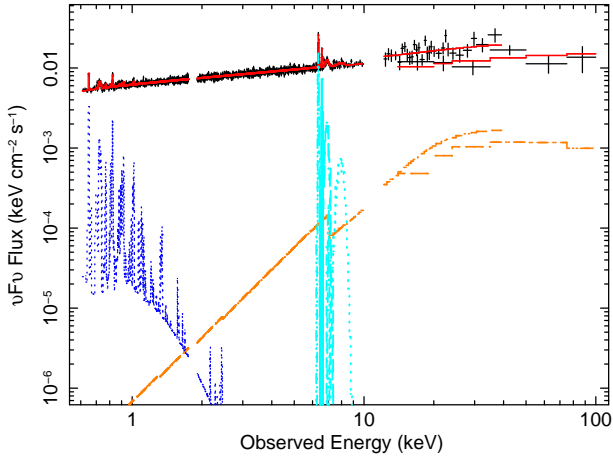}}
\end{center}
\caption{A plot showing the relative contributions of the individual model components across the broad-band 0.6--100\,keV {\sl Suzaku}+BAT energy range (see Section\,3.2 for details). The \textsc{pexrav} neutral reflection component is shown in orange over the 1--100\,keV energy range whilst the soft collisionally-ionised \textsc{mekal} component (at energies \textless 3\,keV) is shown in navy blue. The emission in the Fe\,K complex is modelled with individual Gaussians. The red curve spanning the whole energy range shows the sum of all the model components (including the absorbed power-law continuum) superimposed on the data.}
\end{figure}

\begin{table}
\centering
\begin{tabular}{l c c}
\hline\hline
Model Component & Fit Parameter & Value \\ [0.5ex]
\hline
1. Cutoff Power-law$^{a}$ & $\Gamma$ & $1.74\pm0.01$ \\ & High Energy Cutoff & $> 350$ \\ & Normalisa
tion & $6.28^{+0.05}_{-0.03}$ \\
2. Galactic Absorption$^{b}$ & $N_{\rm H}$ & $1.10\times10^{20}$ \\
3. \textsc{pexrav}$^{c}$ & \textbar$R$\textbar & $0.09^{+0.19}_{-0.08}$ \\
4. \textsc{mekal}$^{d}$ & $k_{\rm B}T$ & $0.27^{+0.05}_{-0.04}$ \\ & Normalisation & $1.27^{+0.59}_{-0.38}$ \\ & Luminosity & $2.14\times10^{40}$ \\
5. Fit Statistics$^{e}$ & $\chi^{2}/{\rm dof}$ & $1100/1041$ \\ & Null Probability & $4.40 \times 10^{-2}$ \\
\hline
\end{tabular}
\caption{Spectral Parameters of continuum fit in the 0.6--150\,keV range (see Section 3.2 for details).
$^{a}$ $\Gamma$, photon index; High energy cutoff in units keV; Normalisation in units $\times10^{-3}$ photons\,cm$^{-2}$\,s$^{-1}$. 
$^{b}$ Local Galactic absorption (at $z=0$), units cm$^{-2}$. 
$^{c}$ \textbar$R$\textbar, reflection scaling factor, where a value $R=1$ corresponds to reflection from neutral material subtending $2\pi$\,sr. 
$^{d}$ $k_{\rm B}T$, plasma temperature in units keV; Normalisation 
(Emission Measure $ = \int n_{e} n_{\rm H} dV$) in units $\times10^{63}$\,cm$^{-3}$.
$^{e}$ Reduced chi-squared ($\chi^{2}/dof$) and null hypothesis probability for spectral fit.}
\end{table}

\subsection{Comparison with Past Observations}

\subsubsection{XMM-Newton \& BeppoSAX}

We tested for any long-term variations in the source by applying our 
best-fit broad-band {\sl Suzaku} model to the 2001 simultaneous {\sl XMM-Newton} 
(30\,ks exposure) 
and {\sl BeppoSAX} PDS (38\,ks exposure) observation (May 2001; see Table\,3), 
as published previously 
by Bianchi et al. (2003, 2004) and Starling et al. (2005). 
The ratio of the {\sl XMM-Newton} EPIC-pn spectrum from 0.3--10\,keV to the best fit 
{\sl Suzaku} model, with continuum parameters described in Table 2, is shown in Figure 9. 
It can be seen that compared to the {\sl Suzaku} XIS spectrum, the pn spectrum 
is steeper, while overall the flux was slightly lower in the {\sl XMM-Newton} data with a value of 
$2.19\times10^{-11}$\,erg\,cm$^{-2}$\,s$^{-1}$ over the 2--10\,keV energy range. We also note
that the flux obtained from the {\sl BeppoSAX} data was lower than that obtained with the {\sl Suzaku}
HXD over the 12--100\,keV range ($2.46 \times 10^{-11}$ compared to $3.81 \times 10^{-11}$\,erg\,cm$^{-2}$\,s$^{-1}$
respectively). Since the photon index of the power-law continuum is quite hard in this source,
this difference could simply arise from small changes in the intrinsic power-law.
The model is generally in good agreement with the data although subtle changes 
in the continuum can be observed, with the spectral curvature being more apparent in the {\sl XMM-Newton} 
data, e.g., with the spectrum being noticeably steeper below 2\,keV, but somewhat flatter 
above 3\,keV. No strong residuals are present in the iron K band, which suggests that the 
iron line emission has remained constant between the 2001 {\sl XMM-Newton} and 2006 {\sl Suzaku} observations.\\

To quantify the changes in the spectrum, the single power-law continuum used to fit 
the {\sl Suzaku} data in Sections 3.1 and 3.2 was replaced with a broken power-law, 
breaking at 2.19$^{+0.34}_{-0.30}$\,keV with $\Gamma$ values of 1.84$\pm0.01$ and 1.71$\pm0.02$ 
below and above this energy break respectively. Furthermore, a slight softening of the 
spectrum below 1\,keV in the {\sl XMM-Newton} data, as suggested by the bump in the residuals 
around 0.9\,keV (which may due to the Ne\,\textsc{ix} triplet or a blend of emission from iron L-shell lines), 
indicates that the single temperature \textsc{mekal} 
component obtained from the {\sl Suzaku} data was not sufficient to model the soft excess. 
Thus a second \textsc{mekal} component was added to the model with a higher temperature of 
$k_{\rm B}T = 0.86^{+0.20}_{-0.14}$\,keV, which significantly improved the fit ($\Delta \chi^{2} = 19.2$)
compared to the model with only a single temperature plasma. The fit parameters of this 
best-fit model to the {\sl XMM-Newton} data are summarised in Table 4.\\ 

For consistency this two temperature \textsc{mekal} model was then applied to the 
2006 {\sl Suzaku} dataset. The temperatures and normalisations of the \textsc{mekal} were
kept fixed at the best-fit values from the {\sl XMM-Newton} data, as an extended diffuse 
collisional plasma 
may not be expected to vary significantly over time (note that if the parameters are allowed to vary, 
the values obtained from {\sl Suzaku} are consistent with the {\sl XMM-Newton} data, within the errors). 
The broken--powerlaw continuum parameterisation 
was also retained from the {\sl XMM-Newton} fit, although the photon indices and normalisations 
were allowed to vary. A comparison of fit parameters for the 
{\sl Suzaku} and {\sl XMM-Newton} observations is shown in Table 4.
No other significant variations were observed between the two observations; 
the Fe\,K line parameters appear to be consistent with constant values (within the errors) for 
the centroid energy and line fluxes, while the 2001 {\sl BeppoSAX} PDS data show 
no evidence for a reflection hump above 10\,keV (with $R < 0.2$; see also Bianchi et al. 2004), consistent with what is 
found by the {\sl Suzaku} HXD. Furthermore, the lower limit on the high energy cutoff value 
is also constrained to
\textgreater 300\,keV (see also Dadina 2008), in good agreement with the HXD and {\it Swift} BAT. 

\subsubsection{Chandra HETG}

The emission line at 6.60\,keV in the {\sl Suzaku} XIS spectrum is found to be rejected at the $>99.9$\% confidence level as arising from the resonance transition of helium-like iron (see Section 3.1 and Figure 5). As a consistency check, we modelled the archival {\sl Chandra} HETG spectrum (using the latest version of the calibration database; v.4.2.2) at Fe\,K with an absorbed power-law and parameterised the emission lines with simple Gaussians consistent with the values found by Bianchi et al. (2008) (also see Table 4) with emission lines at 6.40, 6.72 and 6.99\,keV. Fixing an additional narrow line ($\sigma = 10$\,eV) at 6.60\,keV (as required by {\sl Suzaku}) in the HETG spectrum was not required by the data but resulted in a value for the equivalent width of EW \textless 21.2\,eV, which is consistent with the equivalent width of EW $=24.4\pm8.0$\,eV found in the {\sl Suzaku} XIS. Likewise, including a narrow Gaussian in the {\sl Suzaku} spectrum with the centroid energy fixed at 6.72\,keV (as found by {\sl Chandra}) is also not required by the XIS data but yields an upper limit on the equivalent width of EW \textless 14.3\,eV, again consistent with the equivalent width of EW $=24.0\pm17.0$\,eV found by Bianchi et al. (2008) with the {\sl Chandra} HETG. Therefore, it appears that the Fe\,K parameters in both datasets are consistent with each other with no evidence of variability detected within the errors.

\begin{figure}
\begin{center}
\rotatebox{-90}{\includegraphics[width=6cm]{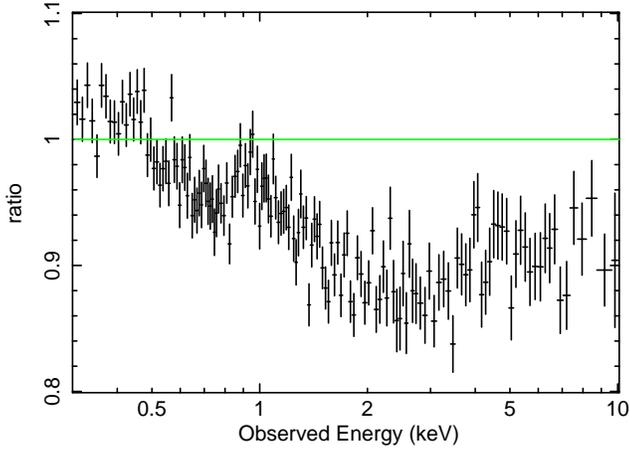}}
\end{center}
\caption{Data/Model residuals of the 2001 {\sl XMM-Newton} EPIC-pn data to the 
{\sl Suzaku} model listed in Table 2. The {\sl XMM-Newton} data show the source to have a 
slightly steeper spectrum compared to {\sl Suzaku} at energies below 2\,keV, while overall the 
continuum flux in the 2--10\,keV band is lower in the {\sl XMM-Newton} data in 2001 compared 
to {\sl Suzaku} in 2006. See Section 3.3 for details.}
\end{figure}

\begin{table}
\centering
\begin{tabular}{l c c c}
\hline\hline
Date & Mission & Instrument & Exposure (ks) \\ [0.5ex]
\hline
27/05/2001 & {\sl BeppoSAX} & MECS & 61 \\
& & PDS & 38 \\
29/05/2001 & {\sl XMM-Newton} & EPIC-PN & 30 \\
15/12/2004 - 27/10/2006 & {\sl Swift} & BAT & 2300 \\
22/10/2006 & {\sl Suzaku} & XIS/HXD & 90 \\
06/08/2007 & {\sl Chandra} & HETG & 148 \\
\hline
\end{tabular}
\caption{Log of observations of NGC 7213. See Section 3.3 for details.}
\end{table}

\begin{table*}
\centering
\begin{tabular}{l c c c c}
\hline\hline
Model Component & Fit Parameter & {\sl Suzaku} & {\sl XMM-Newton} \& {\sl BeppoSAX} & {\sl Chandra}/HETG \\ [0.5ex]
\hline
Flux$^{a}$ & $F_{\rm 0.5-2}$ & 1.37 & 1.29 & - \\
& $F_{\rm 2-10}$ & 2.44 & 2.19 & - \\
& $F_{\rm 12-100}$ & 3.81 & 2.46 & - \\
Broken Powerlaw$^{b}$ & $\Gamma_{1}$ & 1.94$^{+0.09}_{-0.06}$ & 1.84$\pm0.01$ & - \\
& $E_{\rm break}$ & 0.93$^{+0.06}_{-0.07}$ & 2.19$^{+0.34}_{-0.30}$ & - \\
& $\Gamma_{2}$ & 1.74$\pm0.01$ & 1.71$\pm0.02$ & - \\
& Normalisation & 6.11$^{+0.09}_{-0.16}$ & 5.92$^{+0.03}_{-0.04}$ & - \\
\textsc{mekal}$^{c}$ & $k_{\rm B}T_{1}$ & FIXED & 0.24$^{+0.07}_{-0.06}$ & - \\
& Norm$_{1}$ & FIXED & 6.60$^{+4.06}_{-3.81}$ & - \\
& $k_{\rm B}T_{2}$ & FIXED & 0.86$^{+0.20}_{-0.14}$ & - \\
& Norm$_{2}$ & FIXED & 7.63$^{+2.74}_{-2.65}$ & - \\
& $L_{\rm total}$ & 1.99 & 1.99 & - \\
Fe\,K$\alpha$ Line$^{d}$ & $E_{K\alpha}$ & 6.39$\pm0.01$ & 6.40$\pm0.02$ & 6.40$\pm0.01$ \\
& EW$_{\rm K\alpha}$ & 83.1$^{+11.0}_{-10.7}$ & 85.9$^{+14.0}_{-14.4}$ & 120$^{+40}_{-30}$ \\
& Line Flux$_{\rm K\alpha}$ & 2.18$^{+0.28}_{-0.29}$ & 2.33$^{+0.38}_{-0.39}$ & 2.9$^{+0.9}_{-0.7}$ \\
6.60\,keV Line$^{d}$ & $E_{6.60}$ & 6.60$\pm0.03$ & 6.66$^{+0.04}_{-0.05}$ & 6.72$^{+0.01}_{-0.02}$ \\
& EW$_{6.60}$ & 24.1$\pm7.9$ & 27.4$\pm11.6$ & $24\pm17$ \\
& Line Flux$_{6.60}$ & 0.67$\pm0.22$ & 0.78$\pm0.33$ & $0.7\pm0.5$ \\
6.95\,keV Line$^{d}$ & $E_{6.95}$ & 6.95$\pm0.03$ & 6.91$\pm0.08$ & 6.99$^{+0.02}_{-0.01}$ \\
& EW$_{6.95}$ & 62.3$^{+16.0}_{-14.2}$ & 37.1$\pm18.8$ & $60\pm30$ \\
& Line Flux$_{6.95}$ & 1.40$^{+0.36}_{-0.32}$ & 0.89$\pm0.45$ & $1.3\pm0.6$ \\
Statistics & $\chi^{2}$/$d.o.f.$ & $747/723$ & $846/872$ & - \\
& Null Probability & 0.26 & 0.73 & - \\
\hline
\end{tabular}
\caption{Spectral parameters of the broad-band fit in the 0.6--150\,keV energy range for the 2006 {\sl Suzaku} and 2001 simultaneous {\sl XMM-Newton} and {\sl BeppoSAX} observations (see Section 3.3 for details). The Fe-line complex parameters from the 2007 {\sl Chandra}/HETG observation (reported by Bianchi et al. 2008) are also shown for ease of comparison.
$^{a}$ Continuum flux in the specified range in units $\times10^{-11}$\,erg\,cm$^{-2}$\,s$^{-1}$.
$^{b}$ $\Gamma$, photon index; E$_{\rm break}$ in units keV; Normalisation in units $\times10^{-3}$ photons\,cm$^{-2}$\,s$^{-1}$.
$^{c}$ $k_{\rm B}T$, plasma temperature in units keV; Normalisation ($ = \int n_{e} n_{\rm H} dV$) in units $\times10^{62}$\,cm$^{-3}$; Total \textsc{mekal} luminosity from 0.5--10\,keV in units $10^{40}$\,erg\,s$^{-1}$ . All \textsc{mekal} fit parameters in the {\sl Suzaku} dataset are fixed at the best-fit values from the XMM data.
$^{d}$ $E_{\rm line}$, line energy in units keV; EW, equivalent width in units eV; Line Flux in units $\times10^{-5}$ photons\,cm$^{-2}$\,s$^{-1}$.}
\end{table*}

\section{Discussion}

In this section the possible origin of both the neutral and ionised iron K line 
emission from NGC 7213 is discussed along with its implications for the nature of the central engine in this source.

\subsection{The Origin of the Neutral Fe\,K$\alpha$ Line}

We first investigated the possibility of whether a distant Compton-thick reflector, e.g.,
such as the pc-scale torus, could account for the neutral Fe\,K$\alpha$ emission. The tight constraint on the reflection fraction of $R=0.09^{+0.19}_{-0.08}$, obtained in Section 3.2 (also see Table 2), appears
to rule out the possibility of the 6.39\,keV line originating via scattering off Compton-thick
matter since, for an Fe\,K$\alpha$ line with an equivalent width of $\sim$80\,eV (as observed here),
a strong reflection scaling factor value of $R \approx 0.6$ would be required (George \& Fabian 1991).
To test this further, the ionised reflection model \textsc{reflionx} (Ross \& Fabian 2005) 
was used in place of the simple 6.39\,keV Gaussian emission and \textsc{pexrav} 
model for the Compton-scattered continuum off neutral material. The other model 
components, as described in Section 3.2 (also see Tables 1 and 2), were adopted and 
are identical in the spectral fits.
The \textsc{reflionx} model
consists of the emergent spectrum for a photo-ionised optically-thick slab of gas 
when irradiated by a power-law 
spectrum, with a high energy exponential cut-off of 300\,keV, 
using the abundances of Anders \& Ebihara (1982). 
The advantage of the \textsc{reflionx} model is that it 
self-consistently computes both the reflected continuum and 
line emission for the astrophysically abundant elements.\\

We initially fixed the iron abundance to Solar, while the redshift of the reflector 
was found to be consistent with the cosmological redshift of the source, with no
net (e.g., gravitational) redshift. Given the 
narrow unresolved iron K$\alpha$ emission observed in the {\sl Suzaku} spectrum, 
no additional velocity broadening was applied to the reflected spectrum.
We also fixed the reflector ionisation parameter\footnote{Note that in the \textsc{reflionx} ionised reflection model the ionisation parameter is defined as $\xi = \frac{4 \pi F}{n}$ and has units erg\,cm\,s$^{-1}$, where $F$ is the illuminating flux incident upon the reflector 
(integrated over the energy range 100\,eV to 1\,MeV) and 
$n$ is the gas density in cm$^{-3}$.} at a value of $\xi=10$\,erg\,cm\,s$^{-1}$ 
(the lowest value allowed by the model), corresponding to near-neutral iron (i.e.,
iron atoms typically in a low ionisation state corresponding to Fe\,\textsc{i--xvii}). This model provides an upper limit on the reflection scaling factor of $R<0.16$, consistent with what was found by the \textsc{pexrav} model in Section 3, but results in a relatively poor fit to the {\sl Suzaku}+BAT data of 
$\chi^{2}$/$d.o.f.$$=810/715$ (null probability $= 7.22 \times 10^{-3}$). This is due to the fact that 
the model under-predicts the amount of iron K$\alpha$ emission, 
leaving a significant positive residual at $\sim$6.4\,keV in the {\sl Suzaku} XIS data. 
Allowing the Fe abundance to vary to enhance the iron K emission results 
in an acceptable fit of $\chi^{2}$/$d.o.f.$$=745/714$ corresponding to a null hypothesis probability of 0.21 (with $R<0.06$). 
However, in order to adequately model the iron K$\alpha$ line, this  
requires an overabundance of Fe by a factor of $\sim 10$ with respect to Solar (the 90\% confidence lower limit on this value is still 4 times Solar).\\ 

It therefore appears that the lack of an observed Compton reflection hump in the
data above 10\,keV means that the reflection and Fe\,K$\alpha$ emission cannot be
simultaneously modelled in this way, seemingly ruling out a reflection origin for
the Fe\,K$\alpha$ emission, as also suggested by Bianchi et al. (2003) on the basis
of the {\sl BeppoSAX} data. Indeed, an acceptable fit can only be obtained if the
abundances are assumed to be greatly super-Solar, at odds with the modest spread
of values found by Perola et al. (2002) from a sample of nine bright Type 1 Seyferts and NELGs (Narrow Emission-Line Galaxies)
observed with {\sl BeppoSAX}. Thus is appears unlikely that the 6.39\,keV emission originates via reflection off Compton-thick matter unless the material covers
a very small solid angle ($<1$\,sr) and is extremely iron over-abundant. \\

Instead it is perhaps more likely that the near-neutral iron K$\alpha$ line originates 
in Compton-thin matter, covering a higher fraction of $4\pi$ steradians solid angle.
Indeed, an estimate of the column density of the K$\alpha$-emitting material can be made using the calculations of Yaqoob et al. (2010) where an analytic expression relating the efficiency of the Fe\,K$\alpha$ line production and the column density of the emitting material is derived in the optically-thin limit. The production efficiency of the Fe\,K$\alpha$ line is calculated by:

\begin{equation}\chi_{\rm Fe\,K\alpha} = \frac{I_{\rm Fe\,K\alpha, n}}{\int^{\infty}_{E_{\rm K}}E^{-\Gamma}dE}.\end{equation}

Here, $E_{\rm K}$ is the threshold energy for Fe\,K-shell absorption and $I_{\rm Fe\,K\alpha, n}$ refers to the line flux renormalised to an incident continuum with a flux of 1 photon\,cm$^{-2}$\,s$^{-1}$\,keV$^{-1}$ at 1\,keV. $\Gamma$ is the photon index assuming an incident power-law continuum. Adopting the Verner et al. (1996) value for $E_{\rm K}$ of 7.124\,keV, we calculate an Fe\,K$\alpha$ line production efficiency of $\sim$1\% for NGC 7213. In the Compton-thin case, we find that this results in an estimate on the column density of the K$\alpha$-emitting material of $N_{\rm H} \sim 2\times10^{23}$\,cm$^{-2}$ using the analytic expression derived by Yaqoob et al. (2010) (equation 4 in the aforementioned publication). Although this value is consistent with that found by Bianchi et al. (2008), the expression is valid only in the Compton-thin limit which begins to break down for $N_{\rm H} > 2 \times 10^{22}$\,cm$^{-2}$ as the optical depth of the Fe\,K line photons to absorption and scattering becomes non-negligible (see Yaqoob et al. 2010, Figure 2). However, accounting for these effects, according to the calculations of Yaqoob et al. (2010) and Murphy \& Yaqoob (2009), a column density of $N_{\rm H} \sim 3-4 \times10^{23}$\,cm$^{-2}$ can result in an Fe\,K$\alpha$ line efficiency of 1\% for a face-on geometry covering 2$\pi$\,sr in their toroidal X-ray reprocessor model. Thus the Fe\,K$\alpha$ line may originate in a Compton-thin torus or perhaps the outer BLR clouds, as suggested by Bianchi et al. (2005), although the covering fraction would perhaps be slightly high in the latter case (Netzer \& Laor 1993).

\begin{figure}
\begin{center}
\rotatebox{-90}{\includegraphics[width=6cm]{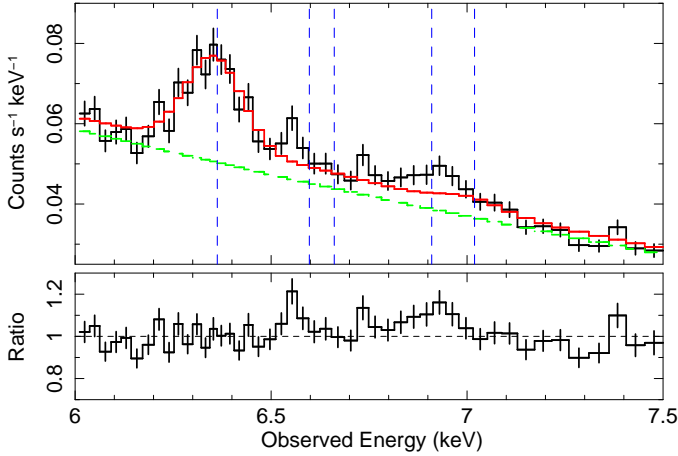}}
\end{center}
\caption{A plot showing the inability to model the highly ionised Fe emission at 6.60\,keV and 6.95\,keV with an ionised reflector therefore suggesting a non-reflection origin for these lines (as described in Section 4.2). The continuum level is shown in green and the lower panel shows the ratio of the residuals to the model. The neutral K$\alpha$ line at 6.39\,keV is modelled with a Gaussian. The vertical dotted lines show the expected line energies of, from left to right, Fe\,K$\alpha$, Fe\,\textsc{xxv} forbidden, Fe\,\textsc{xxv} $1s$--$2p$ resonance, Fe\,\textsc{xxvi} $1s$--$2p$ and Fe\,K$\beta$ in the observed frame.}
\end{figure}

\begin{figure}
\begin{center}
\rotatebox{-90}{\includegraphics[width=6cm]{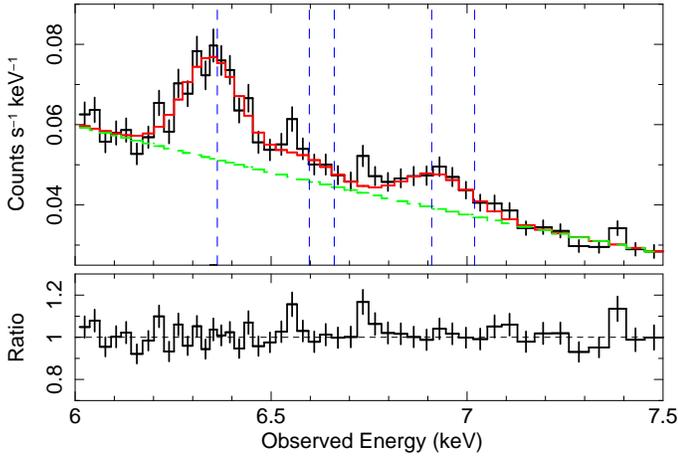}}
\end{center}
\caption{A plot showing the fit obtained when the highly ionised Fe is modelled with a photo-ionised \textsc{xstar} grid with $\sigma = 3000$\,km\,s$^{-1}$ and log\,$\xi = 4.4$\,erg\,cm\,s$^{-1}$ (as described in Section 4.2.2). The continuum level is shown in green and the lower panel shows the ratio of the residuals to the model. The neutral Fe\,K$\alpha$ line at 6.39\,keV is modelled with a Gaussian. The vertical dotted lines show the expected line energies of, from left to right, Fe\,K$\alpha$, Fe\,\textsc{xxv} forbidden, Fe\,\textsc{xxv} $1s$--$2p$ resonance, Fe\,\textsc{xxvi} $1s$--$2p$ and Fe\,K$\beta$ in the observed frame.}
\end{figure}

\subsection{The Origin of the Highly Ionised Fe}

We investigated the possibility of an ionised accretion disc as a potential 
origin for the 6.60 and 6.95\,keV emission lines. To test this scenario, we attempted to model 
the two ionised emission lines by a 
\textsc{reflionx} component with a high ionisation parameter of log\,$\xi = 3.0^{+0.2}_{-0.1}$\,erg\,cm\,s$^{-1}$. 
The remainder of the {\sl Suzaku} spectrum was modelled as before, i.e., 
a cut-off power-law for the continuum emission, a single temperature thermal \textsc{mekal} 
component for the weak soft X-ray excess and a narrow Gaussian centered at 6.39\,keV for the neutral 
iron K$\alpha$ emission. No relativistic blurring was applied. This resulted in a poor fit of $\chi^{2}$/$d.o.f.$$=791/717$ (null probability $= 2.79 \times10^{-2}$) 
with the model unable to account for the ionised emission from either Fe\,\textsc{xxv} 
or Fe\,\textsc{xxvi} (see Figure 10). 
Indeed due to the large intrinsic electron scattering depth at such high values of $\xi$, the 
lines in the reflection model become too broadened and so are unable to model the 
relatively narrow 6.60 and 6.95\,keV 
lines that are observed in the {\sl Suzaku} spectrum. Furthermore the 90\% confidence upper-limit 
on the reflection fraction for the highly ionised reflector is restricted to $R<0.06$, which 
seemingly allows us to reject the presence of a highly ionised Compton-thick medium in NGC 7213, 
such as a highly ionised, but Compton-thick inner accretion disc. \\

In addition, the inability to model the neutral Fe\,K$\alpha$ emission with a low ionisation
reflector (see Section 4.1) coupled with the apparent lack of a Compton hump \textgreater 10\,keV means that
this source exhibits no evidence at all for any Compton-thick material, either neutral or ionised.
Furthermore, the data do not appear to show any relativistic signatures since the neutral and
ionised iron lines appear to be narrow (or only moderately broad in the case of Fe\,\textsc{xxvi}),
thus ruling out any emission from the innermost regions around the black hole. Combining the lack
of evidence for any Compton-thick matter in this source with the lack of any relativistic signature
seems to suggest the complete absence of an inner optically-thick accretion disc in NGC 7213.
Instead it may support the notion of an accretion disc which is truncated at some radius
with the inner regions replaced by some form of Radiatively Inefficient Accretion Flow (RIAF; Narayan \& Yi 1995), 
where the low efficiency of the inner, hot corona leads to much of the energy being 
advected into the black hole rather than radiated away.

\subsubsection{Collisionally-Ionised Model}

We instead investigated the possibility of a 
collisionally-ionised origin for the highly ionised Fe\,K 
emission which we modelled with a high temperature broadened \textsc{apec} 
component (Smith et al. 2001). This required a temperature of $k_{\rm B}T = 13.03^{+2.61}_{-2.86}$\,keV and velocity broadening, $\sigma \sim 2500$\,km\,s$^{-1}$, resulting in a fit statistic of $\chi^{2}$/$d.o.f.$$=788/716$. 
Inspection of the data around the iron K band 
reveals that the collisionally-ionised model over-predicts the contribution of the resonance 
line compared to the forbidden line of Fe\,\textsc{xxv}, which leaves some excess line flux 
near 6.60\,keV unmodelled, although the Fe\,\textsc{xxvi} line is modelled well. Thus, although the collisional model cannot be ruled out with high 
confidence in the present data, given that the energy of the 6.60\,keV line appears 
more consistent with the forbidden rather than resonance transition of Fe\,\textsc{xxv}, an additional photo-ionised component would perhaps be required to model this line (see Section 4.2.2). \\

An estimate on the density of the collisionally-ionised material can be calculated from the normalisation of the \textsc{apec} code, which returns $\int n^{2}\,dV \sim 10^{64}$\,cm$^{-3}$. If the observed broadening of the line is assumed to be due to a Keplerian orbit ($\sigma \sim 0.1$\,keV corresponding to a few $10^{4}$\,km\,s$^{-1}$; see Table 1), we can calculate an estimate on the radius of the emitting material from the central black hole of a few $10^{4}$\,R$_{\rm g}$, where 1\,R$_{\rm g} = 1.5 \times 10^{13}$\,cm for a black hole mass of $10^{8}$\,M$_{\odot}$ (Nelson \& Whittle 1995). Adopting this radius of $R \sim 1.5 \times 10^{17}$\,cm results in an estimate of $n \sim 10^{6}$\,cm$^{-3}$ for the density of the emitting material. \\

If the collisional gas is indeed responsible for some of the highly ionised Fe emission then we note that the conditions of the gas must be different from those in the lower temperature gas used to model the weak soft excess at energies \textless 1\,keV (see Section 3.2). The two components require very different temperatures ($0.27^{+0.05}_{-0.04}$ and $13.03^{+2.61}_{-2.86}$\,keV for the low and high temperature gases respectively) and we note that any contribution from one distinct zone of emission to the other is negligible (i.e., no significant emission from the lower temperature zone is found to be contributing the emission at Fe\,K and vice-versa). However, the two zones of gas could be linked by their origin albeit on completely different scales with the emission from the lower temperature gas originating at a much greater distance from the black hole.

\subsubsection{A Photo-ionisation Model for the Ionised Iron K Emission}

An alternative origin for the lines could be emission from a photo-ionised, but Compton-thin 
plasma. We modelled this scenario using the \textsc{xstar} 2.1ln11 code (Kallman \& McCray 1982), 
which incorporates the abundances of Grevesse, Noels \& Sauval (1996). 
We initially modelled the lines using a single zone of emission with Solar abundances 
and a turbulent velocity width of 200\,km\,s$^{-1}$. 
The best-fit value of the ionisation parameter 
was log\,$\xi = 3.7\pm0.3$ (where the ionisation parameter in \textsc{xstar} 
is defined as in equation 3), which resulted in an acceptable fit with 
$\chi^{2}$/$d.o.f.$$=779/715$ (null probability $= 4.73 \times10^{-2}$). 
Even so, some slight excess residuals were apparent, 
particularly around the 6.95\,keV line, which might suggest there is 
some intrinsic velocity broadening
of the ionised lines. For instance, as measured from Section 3.1, the H-like line appears to have a 
FWHM of $\sim$10\,000\,km\,s$^{-1}$. \\

Thus an alternative \textsc{xstar} grid was used with a 
higher turbulence velocity of $\sigma=3000$\,km\,s$^{-1}$. 
This resulted in a slightly better fit of $\chi^{2}$/$d.o.f.$$=765/715$ (null probability $= 8.50 \times10^{-2}$), 
with an ionisation parameter of log\,$\xi = 4.4\pm0.1$ (see Section 4.3.3 for a calculation of the lower limit on the column density). This fit is significantly better than the one obtained with the collisionally-ionised model (Section 4.2.1) with $\Delta\chi^{2} = 23$ between the two models. A plot of this model superimposed on the data is shown in Figure 11. We also calculated the 90\% uncertainty on the redshift of the zone allowing us to constrain the 
net velocity shift of the ionised emitter to be $v = +650^{+1650}_{-1500}$\,km\,s$^{-1}$, 
where a positive velocity 
denotes redshifted / infalling material. Thus the data do not formally require a velocity shift 
in this model. Further zones of ionised matter are also not required by the data.\\ 

Both high and low turbulence models appear to give good fits over the iron K band, 
with the 6.95\,keV emission 
line originating from the Fe\,\textsc{xxvi} Ly$\alpha$ doublet 
and the 6.60\,keV He-like line arising due to a blend of the forbidden 
and intercombination lines at 6.637 and 6.668\,keV respectively.  
In the low density limit assumed in the \textsc{xstar} model here 
(where $n<10^{16}$\,cm$^{-3}$), the He-like emission is from an approximately equal mixture of the forbidden and intercombination lines, with a negligible 
contribution from the resonance line at 6.700\,keV.
Note that some weak emission via satellite lines of lower ionisation iron 
(i.e., Fe\,\textsc{xxiii-xxiv}) could also be contributing to this 
blend of emission although at the high ionisation parameter inferred here (of $\log \xi 
\sim 4$), this contribution is likely to be negligible (Kallman \& McCray 1982). 
However, future calorimeter-based spectroscopy, e.g., with Astro-H, 
will be required to spectrally resolve all 
the line emission components associated with the He-like triplet of iron and to 
constrain the intrinsic velocity broadening. \\

 We also note that if the highly ionised
iron lines are originating in photo-ionised gas then the possibility remains that a similar
photo-ionised gas but with a much lower ionisation parameter could be responsible for the weak soft
X-ray lines observed at energies \textless 1\,keV. However, Starling et al. (2005) analysed the results of an {\sl XMM-Newton}
RGS observation of NGC 7213 and found that a collisionally-ionised thermal plasma was preferred by the data
from consideration of the '{\sl G}' ratio (Porquet \& Dubau 2000) of the O\,\textsc{vii} triplet.

\subsubsection{The Location of the Highly Ionised Gas}

From consideration of various photo-ionised and collisionally-ionised models, it appears that the 6.60 and 6.95\,keV lines are consistent with originating 
in a photo-ionised medium, although the H-like Ly$\alpha$ line appears to be quite strong 
with an equivalent width of $\sim$60\,eV. Bianchi \& Matt (2002) calculate the equivalent widths 
of Fe\,\textsc{xxv} and Fe\,\textsc{xxvi} lines with respect to both the reflected and total 
continua and show that an equivalent width of $\sim$15\,eV for H-like Fe would be expected if it 
originated in material of column density, $N_{\rm H} \sim 10^{23}$\,cm$^{-2}$, with an ionisation
parameter of log\,$\xi \sim 3.5$, photon index of $\Gamma = 1.7$ and a covering fraction, $f = 1$. 
To test whether our equivalent width of $\sim$60\,eV was feasible we calculated a lower limit 
on the column density of the photo-ionised material in the case of one single zone of emission.\\ 

For a uniform, spherical, Compton-thin shell, 
the normalisation of the photo-ionised emission component is defined within the 
\textsc{xstar} code
by\footnote{see http://heasarc.gsfc.nasa.gov/docs/software/xstar/docs/html/node94.html}:

\begin{equation}k = f_{\rm cov}\frac{L_{\rm ion}}{D^{2}}\end{equation}

where $f_{\rm cov}$ is the covering fraction of the material (i.e., $f_{\rm cov}=1$ for 
matter covering $4\pi$\,sr$^{-1}$), L$_{\rm ion}$ is the ionising 
luminosity in units $10^{38}$\,erg\,s$^{-1}$ from 1 to 1\,000 Rydbergs and $D$ is the 
distance to the source in kpc. Physically the \textsc{xstar} normalisation is simply 
proportional to the observed X-ray flux of the source multiplied by the covering fraction 
of the photo-ionised gas.
Thus the appropriate value of this normalisation, $k$, for NGC 7213 can be calculated assuming 
a covering fraction of 1. For the 
luminosity and distance of NGC\,7213, this results 
in a value of $k = 9.0\times10^{-5}$ in units $10^{38}$\,erg\,s$^{-1}$\,kpc$^{-2}$, where we have adopted a 
distance of 25\,000\,kpc to NGC 7213 (Mould et al. 2000). 
The value used for the ionising 
luminosity was derived from an extrapolation of the broken 
power-law continuum in Table 4, integrated from 1 to 1\,000 Rydbergs 
and is found to be $5.61 \times10^{42}$\,erg\,s$^{-1}$, comparable to the estimate of 
the bolometric luminosity 
of $\sim$$9 \times 10^{42}$\,erg\,s$^{-1}$ by Starling et al. (2005). \\

In the model used in Section 4.2.2, we have fixed the normalisation of 
the additive \textsc{xstar} emission component to the above value and thus 
assumed a fully covering shell of gas around NGC\,7213, with the appropriate source 
luminosity and distance from above. 
The column density was then allowed to vary within the \textsc{xstar} model, in order 
to fit the ionised iron K emission lines. This provided a best fit value for the 
column density of $N_{\rm H}=4.0^{+0.5}_{-0.8} \times 10^{23}$\,cm$^{-2}$. 
The 90\% confidence lower limit on the column density is 
$N_{\rm H}>3.2 \times 10^{23}$\,cm$^{-2}$ for a fully covering spherical 
shell of gas, with an ionisation parameter of $\log \xi \sim 4$ as above. For a shell that does not fully cover the source, then the column density will need to be higher to compensate for the lower covering. This is in good agreement with the calculations of Bianchi \& Matt (2002), who predict 
an equivalent width for Fe\,\textsc{xxvi} Ly$\alpha$ of 15\,eV 
for a column density of $10^{23}$\,cm$^{-2}$, compared to the observed 
60\,eV equivalent width in the case of NGC 7213, but for a column 
density approximately 3--4 times higher. \\

In order to better constrain the origin of the Fe\,\textsc{xxvi} Ly$\alpha$ line it is 
important to estimate a value for the distance of the emitting material. Assuming a uniformly ionised, spherical shell of gas, the ionisation parameter in \textsc{xstar} is defined as:

\begin{equation}\xi = \frac{L_{\rm ion}}{nR^{2}}\end{equation}

and has units erg\,cm\,s$^{-1}$ where $L_{\rm ion}$ is the ionising luminosity from 1 to 1\,000 Rydbergs, $n$ is the gas density in cm$^{-3}$ and $R$ is the radius of the absorbing / emitting material from the central source of X-rays. Combining this with the column density which is given by:

\begin{equation}N_{\rm H} = \int^{\infty}_{R_{\rm in}} n\,dR \end{equation}

yields an estimate on the inner radius of the emitting material:

\begin{equation}R_{\rm in} \sim \frac{L_{\rm ion}}{N_{\rm H} \xi} .\end{equation}

Assuming values of L$_{\rm ion} \sim 5\times10^{42}$\,erg\,s$^{-1}$ (derived above), 
$\xi \sim 5\,000$ and $N_{\rm H} \sim 3 \times10^{23}$\,cm$^{-2}$ then gives a lower limit on 
R of $\sim$$3 \times10^{15}$\,cm corresponding to a value of $\sim$200\,R$_{\rm g}$ 
(assuming a black hole mass of $10^{8}$\,M$_{\odot}$; Nelson \& Whittle 1995) 
and infers an electron density of $n_{e} \sim 10^8$\,cm$^{-3}$ (equation 3). 
This radius is also consistent with the FWHM of $\sim 10\,000$\,km\,s$^{-1}$ 
of the H-like Fe line which, if the broadening is assumed to be the intrinsic 
broadening due to a Keplerian orbit, provides an estimate on the line emitting radius of a few $10^{3}$\,R$_{\rm g}$.

\subsection{NGC 7213 as a Low Luminosity AGN}

\subsubsection{The Inner Advective Flow in NGC 7213}

From 2--10\,keV, the X-ray spectrum of NGC 7213 resembles that of a typical Type 1 Seyfert Galaxy where the spectrum is dominated by a power-law continuum of $\Gamma = 1.75$ and near-neutral Fe\,K$\alpha$ emission at 6.39\,keV. This neutral K$\alpha$ emission may originate from Compton-thin material of $N_{\rm H} \sim 3-4 \times10^{23}$\,cm$^{-2}$ possibly in the outer BLR or a Compton-thin torus (Bianchi et al. 2008). We also detect significant emission from highly ionised material located close to the central source with Fe\,\textsc{xxv} and Fe\,\textsc{xxvi} perhaps originating in a photo-ionised medium with a column density $N_{\rm H}$ \gtsima $3 \times 10^{23}$\,cm$^{-2}$ invoked to match the high observed EW of $\sim$60\,eV. This emission is likely to originate at a distance of $R \sim 10^{3} - 10^{4}$\,R$_{\rm g}$ from the black hole / X-ray source.  \\

Given the tight constraints on reflection from both neutral and ionised material ($R < 0.16$ and $R < 0.06$ respectively), the lack of any relativistic signatures and the very weak Big Blue Bump (Wu et al. 1983; often interpreted as thermal emission from the disc), this appears to rule out the presence of an inner, 'classic' optically thick, geometrically thin accretion disc (Shakura \& Sunyaev 1973) envisaged in the unification scheme of AGN (Antonucci 1993). Instead, we suggest that the accretion disc is maybe truncated at some radius on the order of $10^{3} - 10^{4}$\,R$_{\rm g}$ ($\sim$ 0.01\,pc) with the inner regions perhaps replaced by a Radiatively Inefficient Accretion Flow (RIAF; Narayan \& Yi 1995) consisting of highly ionised, low density ($n \sim 10^{6}$\,cm$^{-3}$), Compton-thin gas covering some significant fraction of $4 \pi$\,sr. In this scenario, the low accretion rate of the source ($0.07\%$\,$L_{\rm Edd}$), perhaps due to a lack of available accreting material, may not allow the infalling material to cool sufficiently in order for a standard thin accretion disc to form. Instead, a stable accretion flow can still occur if the material takes on the form of an optically-thin, hot corona, where most of the material is advected across the event horizon as opposed to radiating away the energy it has acquired in moving close to the black hole. \\

Such a hot, inner flow is expected to comprise of a low-density plasma whereby collisional processes dominate over photo-ionisation due to the high temperature. At radii below about 100\,R$_{\rm g}$, the ion and electron temperatures diverge forming a two-temperature medium with $T_{\rm e} \sim 10^{9} - 10^{10}$\,K and $T_{\rm ion}$ approaching $10^{12}$\,K in the innermost regions (Narayan \& Yi 1995). The electron temperature $T_{\rm e}$ is then expected to fall as $10^{12}\,({\rm K}) / R$ for R $> 10^{2}$, where R is in Schwarzschild units. However, further out at R $\sim 10^{3} - 10^{4}$\,R$_{\rm S}$, a plasma temperature of a few $10^{8}$\,K (i.e., $K_{\rm B}T \sim 13$\,keV, as observed in the collisional \textsc{apec} model), corresponding to $R \sim 10^{4}$\,R$_{\rm S}$, would produce emission from both He-like and H-like Fe, as observed. However, while the Fe\,\textsc{xxvi} emission could plausibly originate in such a plasma, the presence of the Fe\,\textsc{xxv} forbidden line suggests that the He-like Fe cannot be solely produced in such a collisionally-ionised medium.
 Thus an alternative picture could instead be that both the Fe\,\textsc{xxv} and Fe\,\textsc{xxvi} emission lines originate in photo-ionised gas, perhaps at the transition region between the RIAF and the cold, outer accretion disc at a radius $R \sim 10^{3} - 10^{4}$\,R$_{\rm g}$ from the black hole. Alternatively, the Fe\,\textsc{xxv / xxvi} emission lines may arise from a hybrid of photo- and collisionally-ionised processes.

\subsubsection{NGC 7213 as a Low/Hard State Source}

The accretion rate of NGC 7213 is much lower than the predicted ``critical'' value of $\sim$2\%\,$L_{\rm Edd}$ (Maccarone 2003) whereby the high/soft state in X-ray binaries can be observed. As a result, one interesting possibility is that NGC 7213 is an AGN analogue of the low/hard state observed in Galactic Black Hole Candidates (GBHCs). Long-term monitoring of NGC 7213 with RXTE (Phil Uttley, priv. comm.) shows that the AGN is only slowly variable, indicating a relatively low frequency PSD break. This would be consistent with NGC 7213 having a relatively high black hole mass (e.g., M$_{\rm BH} = 10^{8}$\,M$_{\odot}$; Nelson \& Whittle 1995; Woo \& Urry 2002) but a low accretion rate compared to Eddington, consistent with the scaling relations in the timing properties seen between AGN and GBHCs (McHardy et al. 2006). This is further supported by the SED of NGC 7213 which suggests that this object has interesting radio properties lying between those of Radio-Loud and Radio-Quiet Quasars. Indeed, taking the 5\,GHz radio flux and B band flux (host galaxy-subtracted) of Sadler (1984) and Halpern \& Filippenko (1984) respectively and using the equation for radio-loudness\footnote{$R_{\rm L} = {\rm log}_{10}$($F_{\rm 5\,GHz} / F_{\rm B}$), where $F_{\rm 5\,GHz}$ and $F_{\rm B}$ are the 5\,GHz and B band fluxes respectively. Typically, a value of $R_{\rm L} \geq 1$ signifies a Radio-Loud object.} of Wilkes \& Elvis (1986) gives a value of $R_{\rm L} \sim 2$ suggesting that NGC 7213 is intermediate between Radio-Quiet AGN and Radio-Loud AGN such as Radio-Galaxies and Blazars. However, Panessa et al. (2007) also find that a radio-loudness of $\sim$2 is not so uncommon in Seyfert galaxies. \\

The inability to constrain the high energy cutoff of the X-ray emission (i.e., \textgreater 350\,keV) could indicate that the continuum emission may be of non-thermal origin with one possibility being that some of the hard X-ray emission that we observe with {\sl Suzaku} originates from the base of a jet\footnote{Note that the high energy cutoff was measured with {\sl BeppoSAX} in 2001 to be $E_{\rm cut} = 90^{+50}_{-20}$\,keV (Bianchi et al. 2004). However, we find no evidence for a cutoff energy \textless 350\,keV with our combined {\sl Suzaku} XIS, HXD and {\sl Swift} BAT spectra.}. Hameed et al. (2001) imaged NGC 7213 in the optical band and discovered a giant H$\alpha$ filament approximately 19\,kpc from the nucleus. They suggest that such a filament could be the signature of neutral gas shock-ionised by the interactions of a jet. A more recent 8.4\,GHz Long Baseline Array (LBA) radio study of NGC 7213 (Blank et al. 2005) reports that the source is unresolved on the scale of $\sim 3$ milli-arcseconds (corresponding to $\sim$$10^{4}$\,R$_{\rm S}$ at the distance and black hole mass of NGC 7213), just showing a core, indicating that the jet could be orientated face-on. At lower frequencies, there is also evidence for a large-scale structure (30--40$^{\prime\prime}$; Blank, Harnett \& Jones 2005 and reference therein) which could possibly be a signature of the extended radio lobes. Consequently, NGC 7213 is perhaps consistent with the hypothesis of Falcke et al. (1996) whereby the radio-intermediate objects are similar to Radio-Quiet AGN but with moderate beaming from pc-scale jets orientated face-on to the observer.

\subsubsection{The Origin of the High Energy Continuum}

NGC 7213 is not detected to date with the {\sl Fermi} LAT gamma-ray instrument at $\sim$\,GeV energies (Abdo et al. 2010) where the inverse-Compton emission from the jet would be expected to dominate. The {\sl Fermi} LAT flux limit corresponding to the detection threshold of Abdo et al. (2010) at the Galactic co-ordinates of NGC 7213 and for $\Gamma = 1.75$ is $F_{\rm 0.1-100\,GeV} < 3 \times 10^{-9}$\,photons\,cm$^{-2}$\,s$^{-1}$. Extrapolating our best-fit broad-band {\sl Suzaku} model to GeV energies over-predicts the $\gamma$-ray flux by a factor of $\sim\times100$ returning a predicted photon flux of $F_{\rm 0.1-100\,GeV} \sim 3 \times 10^{-7}$\,photons\,cm$^{-2}$\,s$^{-1}$. This implies that the X-ray continuum does in fact roll over at energies \textgreater 350\,keV. \\

In order to be consistent with the 0.1--100\,GeV flux limit from {\sl Fermi}, we require that the E-folding energy of the power-law component must be $E_{\rm cut} < 25$\,MeV. This, combined with the lower limit on the high-energy cutoff from the combined {\sl Suzaku} / {\sl Swift} data means that 350\,keV \textless $E_{\rm cut}$ \textless 25\,MeV, consistent with the predicted electron temperature of the hot, inner flow (see Section 4.3.1). This suggests that thermal Comptonisation is responsible for the X-ray continuum and that any non-thermal contribution from the inverse-Compton component associated with a jet may be negligible in this source. \\

 Furthermore, the high EW of the observed emission lines may also suggest that there is very little dilution of the X-ray continuum by a jet. We note that other radio-loud sources such as the Broad-Line Radio Galaxies (BLRG) 3C 120 and 3C 390.3 do also show fairly strong Fe\,K line emission with EWs on the order of 50--100\,eV. However, in the case of 3C 390.3, Sambruna et al. (2009) argue from the {\sl Suzaku} data and the overall radio - $\gamma$-ray SED that the jet makes a minimal contribution to the X-ray continuum emission. Likewise, Kataoka et al. (2007) argue a similar case from the {\sl Suzaku} observation of 3C 120 and conclude that the putative jet component does not dilute the Fe\,K emission. In addition, comparing the ratio of the 1--100\,GeV $\gamma$-ray flux from {\sl Fermi} to the 2--10\,keV X-ray flux for NGC 7213 with that of 3C 111 (the only BLRG detected by {\sl Fermi} to date) and 3C 273, we find that the ratio is higher by a factor of $>\times6$ for the two radio-loud AGN. By comparison, like for NGC 7213, none of the X-ray bright Type 1 Seyferts appear to have been detected with {\sl Fermi} thus far. \\

As a further test, we did attempt to model the X-ray continuum with a double power-law component consisting of a hard spectrum to model any possible emission from a jet and a much softer, Seyfert-like spectrum to model the photo-ionising nuclear X-ray emission. Statistically speaking, this fit is not required by the data as it only yields an improvement of $\Delta \chi^{2} \sim 2$ for an additional two parameters of interest. Upon fixing the photon index of the softer power-law at $\Gamma = 2$, we find that the normalisation of this component becomes very small with an upper-limit corresponding to just 10\% of the normalisation of the main power-law. So it seems that a double power-law model is not required by the data and that the X-ray continuum is best represented by a single power-law component. \\

The photon index of the X-ray continuum has a best-fit value of $\Gamma = 1.75\pm0.02$ and therefore has only a slightly flatter spectrum than the typical values usually associated with Radio-Quiet Quasars (RQQs) and Type 1 Seyferts. For example, Reeves \& Turner (2000) find a mean value of $\Gamma = 1.89\pm0.05$ from a sample of 27 RQQs observed with {\sl ASCA} (Advanced Satellite for Cosmology and Astrophysics) and Nandra \& Pounds (1994) find a mean value of $\Gamma = 1.95$ with a dispersion of $\sigma = 0.15$ from their sample of Seyfert galaxies observed with {\sl Ginga}. Porquet et al. (2004) also find a mean value of $\Gamma = 1.90$ with a dispersion of $\sigma = 0.27$ from a sample of 14 RQQs observed with {\sl XMM-Newton}. The photon index of NGC 7213 is, however, consistent with those found in other low-luminosity AGN such as M81 (Young et al. 2007) and NGC 4579 (Dewangan et al. 2004). Interestingly, the spectrum in NGC 7213 does appear to be somewhat steeper than the predicted photon index of $\Gamma = 1.4$ from the relation between mass accretion rate and photon index of Papadakis et al. (2009) given its calculated accretion rate of 0.07\%\,$L_{\rm EDD}$ (an accretion rate of $\sim$2\%\,$L_{\rm EDD}$ would be required to obtain $\Gamma = 1.75$). So it seems that NGC 7213 may not strictly follow the positive correlation between spectral steepness and accretion rate in AGN and X-ray Binaries suggested by Shemmer et al. (2006) although the high energy cutoff of the X-ray continuum (i.e., 350\,keV \textless $E_{\rm cut}$ \textless 25\,MeV) observed here may suggest that the hard X-ray emission that we observe with {\sl Suzaku} and {\sl Swift} could be consistent with a very hot inner-flow, compatible with NGC 7213 having a low mass accretion rate.

\section{Conclusions}

NGC 7213 is an unusual AGN as it consistently exhibits no evidence for a Compton reflection component unlike in other Type 1 Seyferts (Perola et al. 2002; Dadina 2008). The time-averaged continuum emission is well fitted by a single power-law of $\Gamma = 1.75$ and from consideration of the combined {\sl Suzaku} and {\sl Swift} BAT data and the {\sl Fermi} flux limit we constrain the high energy cutoff to be 350\,keV \textless $E_{\rm cut}$ \textless 25\,MeV. The limits on reflection for the neutral and ionised cases from the \textsc{reflionx} model are  $R < 0.16$ and $R < 0.06$ respectively suggesting that a significant Compton-thick reflector (e.g., from the inner disc or Compton-thick torus) is absent in this source, consistent with previous findings (e.g., Bianchi et al. 2003). Nonetheless, a significant Fe\,K complex is observed above 6\,keV appearing only in emission. The line from neutral K$\alpha$ dominates (6.39\,keV; EW $\sim 80$\,eV) with further contributions from Fe\,\textsc{xxv} and Fe\,\textsc{xxvi}\,Ly$\alpha$ (6.60 and 6.95\,keV respectively; also see Starling et al. 2005, Bianchi et al. 2008). Furthermore, in this observation we also find that the Fe\,\textsc{xxvi}\,Ly$\alpha$ emission appears to be somewhat resolved in the {\sl Suzaku} spectrum with a FWHM $\sim 10\,000$\,km\,s$^{-1}$ and that the emission from Fe\,\textsc{xxv} appears to be consistent with the forbidden transition from helium-like iron at $\sim$6.64\,keV as opposed to the resonance transition at $\sim$6.70\,keV. \\

The neutral K$\alpha$ emission cannot be modelled via reflection off Compton-thick matter. However, an origin in a Compton-thin plasma covering a significant fraction of $4\pi$\,sr is feasible with an inferred column density of $N_{\rm H} \sim 3-4 \times10^{23}$\,cm$^{-2}$, again consistent with the findings of previous observations with {\sl Chandra} and {\sl XMM-Newton}. Likewise here, the emission from highly ionised iron can also be modelled with a substantial column ($N_{\rm H}$ \gtsima $3 \times 10^{23}$\,cm$^{-2}$) of photo-ionised matter if a location close to the central engine is invoked to explain the inherent broadening and the high ionisation state. Given the lack of either neutral or ionised reflection coupled with the apparent absence of any relativistic signature in the spectrum, it appears that an inner, optically-thick accretion disc may be absent in this source. Instead, the accretion disc in NGC 7213 is most likely truncated at some radius with the inner regions perhaps replaced by an advective accretion flow (e.g., RIAF; Narayan \& Yi 1995). The Fe\,\textsc{xxv / xxvi} emission could then be the ionised signature of such a hot, optically-thin plasma originating in material a few $10^{3}$\,R$_{\rm g}$ from the central X-ray source.

\section{Acknowledgements}

This research has made use of the NASA Astronomical Data System (ADS), the NASA Extragalactic Database (NED) and data obtained from the {\sl Suzaku} satellite, a collaborative mission between the space agencies of Japan (JAXA) and the USA (NASA). We wish to thank our anonymous referee for their useful comments and thorough review of the draft. Valentina Braito would also like to acknowledge support from the UK STFC research council.

\end{document}